# Topological and nodal superconductor kagome magnesium triboride

Yipeng An,[1,*] Juncai Chen,[1] Zhengxuan Wang,[1] Jie Li,[2] Shijing Gong,[3] Chunlan Ma,[4] Tianxing Wang,[1] Zhaoyong Jiao,[1] Ruqian Wu,[2,†] Jiangping Hu,[5,6,‡] and Wuming Liu[5,§]

[1]School of Physics, Henan Normal University, Xinxiang 453007, China

[2]Department of Physics and Astronomy, University of California, Irvine 92697, USA

[3]Department of Physics, East China Normal University, Shanghai 200062, China

[4]School of Physics and Technology, Suzhou University of Science and Technology, Suzhou 215009, China

[5]Beijing National Laboratory for Condensed Matter Physics, Institute of Physics, Chinese Academy of Sciences, Beijing 100190, China

[6]Kavli Institute of Theoretical Sciences, University of Chinese Academy of Sciences, Beijing 100190, China

Recently the kagome compounds have inspired enormous interest and made some great progress such as in the field of superconductivity and topology. Here we predict a new kagome magnesium triboride (MgB$_3$) superconductor with a calculated $T_c$ ~12.2 K and $T_c$ ~15.4 K by external stress, the potentially highest among the reported diverse kagome-type superconductors. We reveal its various exotic physical properties including the van Hove singularity, flat-band, multiple Dirac points, and nontrivial topology. The system can be described by a two-band model with highly anisotropic superconducting gaps on Fermi surfaces. Its topological and nodal superconducting nature is unveiled by a recently developed symmetry indicators method. Our results suggest that MgB$_3$ can be a new platform to study exotic physics in the kagome structure, and pave a way to seek for more superconductors and topological materials with XY$_3$-type kagome lattice.

## I. INTRODUCTION

In recent years, the electronic physics in kagome lattices is a new research hotspot as it displays many remarkable physical properties, including flat-band [1], superconductivity [2], nontrivial topology structure [3], charge density wave (CDW) [4], pair density wave (PDW) [5], nematicity [6], anomalous Hall effect [7], and time-reversal symmetry-breaking charge order [8]. These exotic properties have either been theoretically predicted or experimentally observed in diverse kagome layers, including Fe$_3$Sn$_2$ compound (non-superconducting) [9], CoSn-type family ($T_c$ < 1.9 K) [10,11], AV$_3$Sb$_5$ family (0.9 K < $T_c$ < 2.8 K) [2,5,12-18], AV$_6$Sb$_6$ family (non-superconducting) [19], and LaRu$_3$Si$_2$ compound ($T_c$ < 7 K) [20]. Most of these kagome materials are heavy and have low superconducting transition temperatures (i.e., $T_c$ < 7 K).

Magnesium diboride (MgB$_2$), as a phonon-mediated type-II BCS (Bardeen, Cooper, and Schrieffer) superconductor, has been a historically important superconductor for both fundamental research and applications due to its usually high $T_c$ (39 K) and simple structure [21]. Various methods have been attempted to further tune its superconductivities, such as doping electrons or holes [22] and applying pressure and stretch strain [23,24]. The $T_c$ has been accurately calculated (39.3 K) recently by means of *ab initio* calculations with proper functionals [25], i.e., the density functional theory for superconductors (SCDFT) [26-30] which has been used to reproduce the experimental $T_c$ of various superconductors [30-33].

---

*ypan@htu.edu.cn
†wur@uci.edu
‡jphu@iphy.ac.cn
§wliu@iphy.ac.cn

In this paper, we theoretically report a stable magnesium triboride MgB$_3$ (P6/mmm) kagome structure which is comprised of kagome B layer and rhombic Mg layer with Mg embedded in the midpoint of two hexagons of adjacent B layers [Fig. 1(a)]. We identify various exotic physical properties in this structure, including van Hove singularity (VHS), flat-band (FB), multiple Dirac points (MDP) and nontrivial topology. The calculated superconducting transition temperature $T_c$ is the *highest* among the known diverse kagome-type superconductors. In addition, we calculate a pressure-dependent phase diagram of superconducting, metallic, and unstable states and predict that the maximum $T_c$ ~15.4 K under pressure. The topological and nodal superconducting nature of MgB$_3$ is unveiled by a recently developed symmetry indicators method. We also generate scanning tunneling microscopy (STM) image and angle-resolved photoemission spectroscopy (ARPES) picture for the experimental verification with the surface Green's function method.

## II. METHODS

The *ab initio* calculations are performed with the Quantum ESPRESSO (QE) integrated suite [37-39] unless stated otherwise. The SG15 optimized norm-conserving Vanderbilt (ONCV) pseudopotentials [40-42] are employed to describe the effect of core electrons. The generalized gradient approximation with the van der Waals corrected OptB88-vdW functional is adopted for this layered kagome compound MgB$_3$ [43-47]. A 12 × 12 × 12 Monkhorst–Pack $k$-point grid is adopted for the self-consistent calculations, and a denser grid (24 × 24 × 24) is used to obtain the density of states and band structures, while coarse grid (6 × 6 × 6) is adopted for the phonon calculations. The plane-wave kinetic-energy cutoff and the energy cutoff for charge density are set to 80 and 320 Ry, respectively. The total energy tolerance and residual force on each atom are < 10$^{-10}$ Ry and 10$^{-8}$ Ry Bohr$^{-1}$ in the geometry optimization. The phonon frequencies are obtained using the density functional perturbation theory [48]. The optimized tetrahedron method [49] is used for the Brillouin-zone integration to obtain the electron-phonon interaction.

The nontrivial topology properties are obtained by the wannier90 [35] and WannierTools [36] codes including the spin-orbital coupling (SOC) effect. The superconductivity calculations are performed using the SCDFT by the Superconducting-Toolkit [31,32]. The temperature-dependent structure (3 × 3 × 3 supercell) is obtained *via* the special displacement method as implemented in the ZG code [50,51]. The temperature-dependent electronic spectral function is obtained by unfolding the band of the supercell. Their electronic structures of ground states are read from the self-consistent results of QE, which is also used to stimulate the STM picture. The work function (Φ) and APRES pictures (i.e., surface band) of surface system are obtained by the surface Green's function method as implemented in the QuantumATK [52-55].

In the EPC calculations, the mode-resolved EPC $\lambda_{qv}$ is described as

$$\lambda_{\mathbf{q}\nu} = \frac{\gamma_{\mathbf{q}\nu}}{\pi \hbar N(\varepsilon_F)\omega_{\mathbf{q}\nu}^2}, \quad (1)$$

where $N(\varepsilon_F)$ is the DOS at the Fermi level ($\varepsilon_F$), $\omega_{\mathbf{q}\nu}$ refers to the phonon frequency, and $\gamma_{\mathbf{q}\nu}$ is the phonon linewidth obtained by

$$\gamma_{\mathbf{q}\nu} = 2\pi\omega_{\mathbf{q}\nu} \sum_{ij} \int \frac{d^3k}{\Omega_{BZ}} |g_{\mathbf{q}\nu}(\mathbf{k},i,j)|^2 \\ \delta(\varepsilon_{\mathbf{q},i} - \varepsilon_F)\delta(\varepsilon_{\mathbf{k+q},j} - \varepsilon_F) \quad (2)$$

where $g$ is the electron-phonon coefficient, $\varepsilon_{\mathbf{q},i}$ ($\varepsilon_{\mathbf{k+q},j}$) are the Kohn-Sham eigenvalues, $\Omega_{BZ}$ refers to the volume of the Brillouin-zone. The Eliashberg electron-phonon spectral function $\alpha^2F(\omega)$ is obtained by

$$\alpha^2F(\omega) = \frac{1}{2\pi N(\varepsilon_F)} \sum_{\mathbf{q}\nu} \delta(\omega - \omega_{\mathbf{q}\nu}) \frac{\gamma_{\mathbf{q}\nu}}{\hbar\omega_{\mathbf{q}\nu}}. \quad (3)$$

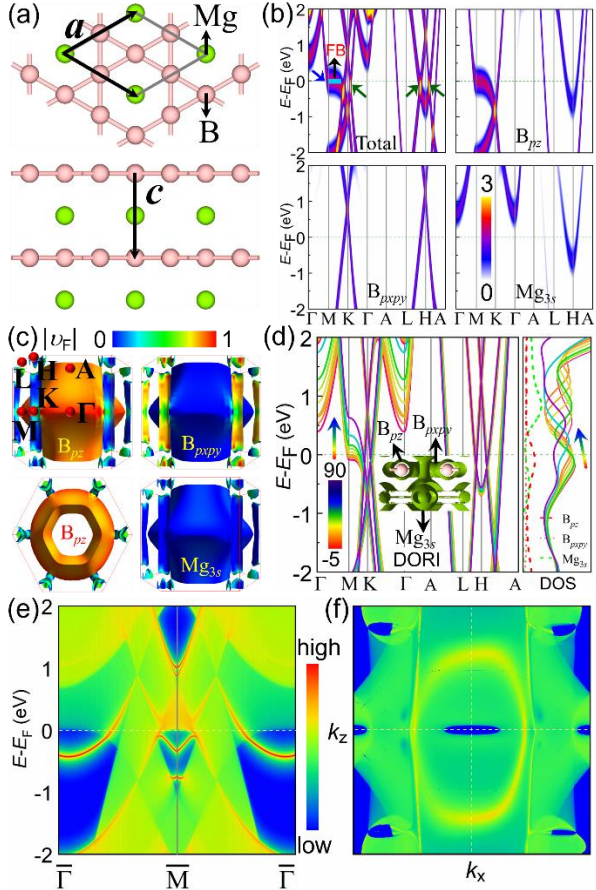

FIG. 1. (a) Top and side views of MgB$_3$ with lattice parameters *a* (*c*) = 3.465(3.593) Å. (b) Projected band and projections of B$_{pz}$, B$_{pxpy}$, and Mg$_{3s}$ orbitals. The blue and green arrows refer to the van Hove singularity and multiple Dirac points, respectively. The skyblue short-line indicates the flat-band (FB) region. (c) Projections of Fermi velocity $|v_F|$ on Fermi surfaces drawn by using the FermiSurfer code [34]. (d) Band and density of states (DOS) under various pressures from −5 to 90 GPa and the projected DOS at zero pressure. The embedded figure in (d) is the Density Overlap Regions Indicator (DORI). (e) Topological surface states projected on semi-infinite (010) surface and (f) Fermi arc (closed orange ellipse) at $E_F$ obtained by wannier90 [35] and WannierTools [36] codes. The Fermi level is set to zero.

The electron-phonon mass enhancement parameter $\lambda$ can be obtained directly from the integration of the $\lambda_{\mathbf{q}\nu}$ in the first Brillouin-zone for all phonon modes or defined as the first reciprocal momentum of the spectral function,

$$\lambda(\omega) = \sum_{\mathbf{q}\nu} \lambda_{\mathbf{q}\nu} = 2\int_0^\omega \frac{\alpha^2F(\omega)}{\omega} d\omega. \quad (4)$$

In the SCDFT calculations, the superconducting gap at a wave-vector $\mathbf{k}$ is described as

$$\Delta_{n\mathbf{k}} = -\frac{1}{2} \sum_{n'\mathbf{k}'} \frac{K_{n\mathbf{k}n'\mathbf{k}'}(\varepsilon_{n\mathbf{k}}, \varepsilon_{n'\mathbf{k}'})}{1 + Z_{n\mathbf{k}}(\varepsilon_{n\mathbf{k}})} \times \\ \frac{\Delta_{n'\mathbf{k}'}}{\sqrt{\varepsilon_{n'\mathbf{k}'}^2 + \Delta_{n'\mathbf{k}'}^2}} \tanh(\frac{\sqrt{\varepsilon_{n'\mathbf{k}'}^2 + \Delta_{n'\mathbf{k}'}^2}}{2T}) \quad (5)$$

where $\varepsilon_{n\mathbf{k}}$ is the $n$th eigenvalue of normal-state Kohn-Sham (KS) orbital measured from the Fermi level. $K_{n\mathbf{k}n'\mathbf{k}'}(\varepsilon, \varepsilon')$ refers to the superconducting-pair creation and annihilation interactions. $Z_{n\mathbf{k}}(\varepsilon)$ is the electron-phonon renormalization factor. $T$ is the temperature defined by setting the Boltzmann constant $k_B = 1$. The superconducting critical temperature $T_c$ is determined using the bisection method [32]. Namely, the initial lower limit $T_c^{min}$ is set to zero, and the initial upper limit $T_c^{max}$ is set according to the BCS theory ($2\Delta_0/3.54$, where $\Delta_0$ is the superconducting gap averaged over Fermi surfaces at 0 K, this value is doubled when there is a finite gap at this temperature). The gap equation 5 is obtained at $T = (T_c^{min} + T_c^{max})/2$, and, either $T_c^{min}$ or $T_c^{max}$ is replaced by $T$ in the following steps based on whether the amplitude of average gap is small enough (i.e., $\langle|\Delta|\rangle < 10^{-3}\Delta_0$). This procedure is repeated 10 times, and $T_c$ is obtained as the average of $T_c^{min}$ and $T_c^{max}$, which are very close to each other.

The topological and nodal superconductivities of kagome MgB$_3$ is examined by a recently established method of symmetry indicators (SIs) [56], and their diagnosis is performed by the recently developed QEIRREPS [57] and Topological Supercon [58] tools. This method has been well used to study the topological and nodal superconductors [59-61].

To simulate the ARPES (i.e., surface band) picture, the spectral function (or local density of states) is described by

$$A(\varepsilon, \mathbf{k}_\parallel) = -\frac{1}{\pi} \text{Im}[\mathbf{G}_{\mathbf{k}_\parallel}(\varepsilon)], \quad (6)$$

$$\mathbf{G}_{\mathbf{k}_\parallel}(\varepsilon) = [\varepsilon - \mathbf{H}_{K_\parallel}^{DFT} - \mathbf{\Sigma}_{K_\parallel}(\varepsilon)]^{-1}, \quad (7)$$

where $\mathbf{k}_\parallel = (k_a, k_b)$ is the transverse $k$-vector, $\mathbf{G}_{\mathbf{k}_\parallel}(E)$ refers to the retarded Green's function of the surface system, $\mathbf{H}_{K_\parallel}^{DFT}$ indicates the DFT Hamiltonian of the surface region and the screening layers, and $\mathbf{\Sigma}_{K_\parallel}(\varepsilon)$ is the self-energy of the semi-infinite bulk part.

### III. RESULTS AND DISCUSSION

#### A. Electronic structures and topology of kagome MgB$_3$

The MgB$_3$ kagome lattice has the space group $P6/mmm$ (No. 191) and point group $D_{6h}$, and its lattice parameters are $a$ ($c$) = 3.465(3.593) Å, respectively. Its formation energy ($E_{fe}$) is a negative value (−2.16 eV), which is obtained by the equation $E_{fe} = (E_{MgB3} − E_{Mg} − 3 × E_B)/4$, where $E_{MgB3}$ is the total energy of MgB$_3$, $E_{Mg/B}$ refers to the elemental energy of Mg crystal and B cluster. The MgB$_3$ can be prepared by the MgB$_2$ and B powder at a special condition because of its negative formation energy (−1.85 eV) given by $E_{MgB3} − (E_{MgB2} + E_B)$. These demonstrate the energy stability of MgB$_3$.

There are several important features in the electronic structure of MgB$_3$. First, as shown in Fig. 1(b), there is van Hove singularity (VHS) in the band structure at the M point near the Fermi level ($E_F$). The presence of the VHS can drive some unusual physical phenomena, such as Fermi surface (FS) instability [62,63]. This VHS moves up or down with external strain [Fig. 1(d)], and DOS peak near $E_F$ shifts up and down accordingly. These intrinsic VHS and FB properties stem from the B$_{pz}$ orbitals of the kagome lattice [see the projections in Fig. 1(b)] which generate the main contributions to the density of states near $E_F$ [see their projections in Fig. 1(d)]. In other words, they constitute a so-called higher-order VHS [64,65] and exhibit less pronounced FS nesting with large DOS [Figs. 1(c) and 1(d)]. The appearance of higher-order VHS has been proposed to be a possible explanation for a CDW instability of kagome lattice [66] and a nematic order [64,65].

Second, multiple Dirac points are observed in the electronic band [Fig. 1(b)], three (see the three green arrows) of which are close to $E_F$ and away from the high-symmetry points of the Brillouin-zone. This suggests that its electronic properties and superconductivities can be tuned by structure strain or doping to move these three Dirac points to $E_F$ [18,67]. The MDP property is ascribed to the mixture of B$_{pz}$, B$_{pxpy}$, and Mg$_{3s}$ orbitals [see their projections in Fig. 1(b)]. We draw the distributions of Fermi velocity $|v_F|$ of these three orbitals on Fermi surfaces with a color map [Fig. 1(c)]. They vary largely over FS and show the trend that B$_{pz}$ > B$_{pxpy}$ > Mg$_{3s}$. The ratio of the maximum to the minimum value of $|v_F|$ is 45. The in-plane B$_{pxpy}$ orbitals form the σ-bonding, while B$_{pz}$ and Mg$_{3s}$ orbitals mix into the π-bonding. They are clearly described by the Density Overlap Regions Indicator (DORI) [68], as shown in the embedded figure in Fig. 1(d). Note these σ- and π-bonding states were observed to dominate the two superconducting gaps of MgB$_2$.

Third, a flat-band region can also be observed along the M-K path (near the $E_F$), which is known to be the essential feature in the magic-angle bilayer graphene systems [69,70].

Finally, the nontrivial topology of time-reversal invariant MgB$_3$ is also observed *via* its nontrivial topological invariants, surface states, and closed elliptic Fermi arc at the $E_F$ [see Figs. 1(e) and 1(f)].

### B. Phonons and electron-phonon coupling

These remarkable electronic features of the MgB$_3$ kagome lattice inspire us to study further its phonons, electron-phonon coupling (EPC), and phonon-mediated superconducting properties. At static zero-pressure, the MgB$_3$ is dynamically stable with the absence of imaginary frequencies in the phonon dispersion [Fig. 2(a)]. The contribution of B vibrations to the phonon dispersion is larger than that of Mg atoms at the low-frequency region below 18 meV, where a strong EPC $\lambda_{\mathbf{q}v}$ (at wavevector **q** and branch $v$) appears and is mainly distributed on the in-plane longitudinal acoustic (LA) branch [see its vibration mode in Fig. 2(b)]. Although B vibrations have a lower weight in phonon bands from 18 to 37 meV, they contribute in the high-frequency region beyond 50 meV due to the smaller mass of B atoms and have large phonon linewidth $\gamma_{\mathbf{q}v}$.

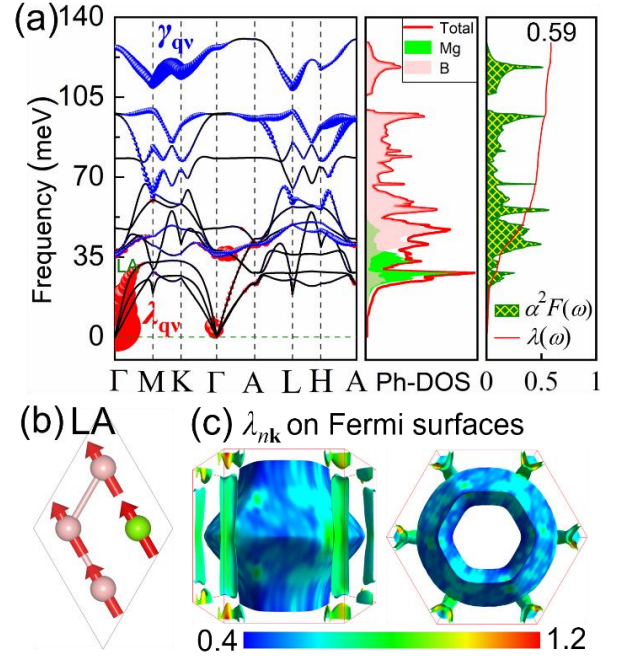

FIG. 2. (a) Phonon dispersions with electron-phonon coupling (EPC) $\lambda_{\mathbf{q}v}$ (red balls) and phonon linewidth $\gamma_{\mathbf{q}v}$ (blue balls), projected phonon density of states (Ph-DOS), and Eliashberg spectral functions $\alpha^2 F(\omega)$ with the cumulative EPC strength $\lambda(\omega)$. (b) Vibration mode of LA branch. (c) Side and top views of EPC $\lambda_{n\mathbf{k}}$ on Fermi surfaces.

The Eliashberg electron-phonon spectral function $\alpha^2 F(\omega)$ [see the right panel of Fig. 2(a)] has many sharp peaks, which stem from the multi-peak phonon density of state (Ph-DOS) and multiple flat-band phonon dispersion. A broad peak of $\alpha^2 F(\omega)$ (i.e., from 37 to 50 meV) leads to the significant increase of the cumulative frequency-dependent EPC strength $\lambda(\omega) = 0.59$, with an average phonon frequency $\omega_{\ln} = 45$ meV (527 K). Phonons in the region below 50 meV contain vibrations from all B and Mg atoms, contributing 57% of total EPC. The EPC of MgB$_3$ has a large magnitude ($\lambda_{\mathbf{q}\nu}$) in the low-frequency region (i.e., below 24 meV), and has a strong anisotropy. The anisotropic ratio (i.e., the maximum to minimum value) of the EPC $\lambda_{n\mathbf{k}}$ (band index $n$ and wave number $\mathbf{k}$ dependent) projected on the Fermi surfaces [Fig. 2(c)] is 2.7 obtained from the SCDFT method. The average value of EPC $\lambda_{\text{avg}}$ is 0.63 (< 1), slightly lower than that of $MgB_2$ (0.80) [25].

## C. Superconductivities and quantum control

The calculated superconducting gap $\Delta_{n\mathbf{k}}$ is also anisotropic (e.g., with a ratio of 1.7 at 0.1 K) [Fig. 3(a)] due to its anisotropic EPC nature, and the average gap value $\Delta_{\text{avg}}$ is 1.9 meV. Its anisotropic ratio and average value are up-and-down when a negative (i.e., stretch strain) or positive hydrostatic pressure is applied. For instance, the anisotropic ratio and average value are 2.4 and 2.5 meV at −5 GPa respectively [Fig. 3(b)]. A quantum phase transition can emerge when the external stress or strain is out of the range from −5 to 90 GPa. An instability phase is observed because some imaginary frequencies appear such as at −5.5 or 100 GPa [see Fig. S2 of the Supplemental Material [71]]. Similar instability phases have also been observed in pristine kagome or hexagonal systems [3,13,72]. This may be ascribed to the shift of the VHS where it is far away from the $E_F$ [Fig. 1(d)]. Note the $\Delta_{n\mathbf{k}}$ of $MgB_3$ has a $s$-wave symmetry with the same sign and is non-zero everywhere on FS, suggesting its type-II full gap $s$-wave superconductor type. In real experimental setup, the pressure can be realized such as using the miniature diamond anvil cell by using NaCl as a pressure transmitting medium [73]. The pressure can be calibrated by using the shift of ruby florescence and diamond anvil Raman.

At static zero pressure, the intrinsic $T_c$ of $MgB_3$ superconductor is predicted to be 12.2 K, the *highest* among the reported kagome lattice systems. It can be raised to 15.4 K when a stretch strain of −5 GPa is applied, and falls to a saturation value of 5.9 K with the further increase of pressure. The mechanism of pressure-dependent $T_c$ can be easily understood such that the increase of pressure leads to the decrease of lattice [see Fig. S3 of the Supplemental Material [71]], DOS at $E_F$ $N(\varepsilon_F)$, EPC $\lambda$, $\Delta$, and $T_c$. Based on these findings, we plot its pressure-dependent $T_c$ and $\Delta_{\text{avg}}$, as well as the quantum phase diagram of superconducting, metallic, and unstable states in Fig. 3(c). The electron- and hole-doping can also be expected to tune $T_c$ [22,74].

Figure 3(d) shows the calculated pressure-dependent normalized superconducting quasiparticle density of states (QPDOS), which can be used to compare directly with the experimental tunneling conductance. One can find that the $MgB_3$ superconductor has two superconducting gaps $\Delta_\sigma$ and $\Delta_\pi$, which are induced by its $\sigma$- and $\pi$-bonding states respectively. Their amplitudes decrease and their positions are gradually shifted to the low-energy region with the increase of pressure. Its superconducting gaps also display reasonable attenuation and shift in the temperature-dependent QPDOS [see Fig. S5 of the Supplemental Material [71]].

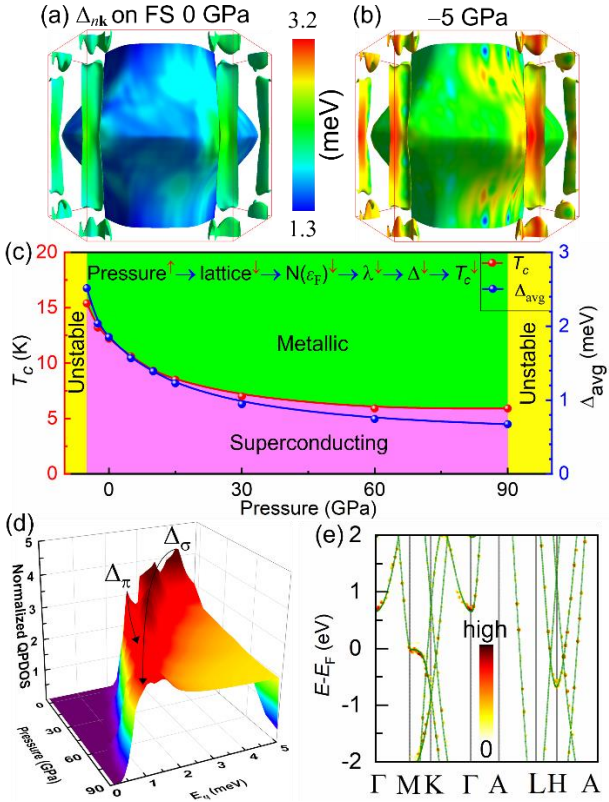

FIG. 3. Superconducting gap $\Delta_{n\mathbf{k}}$ on Fermi surfaces at 0.1 K under the pressures of 0 (a) and −5 GPa (b). (c) Pressure-dependent $T_c$ and $\Delta_{\text{avg}}$, as well as the quantum phase diagram of superconducting, metallic, and unstable states. The embedded illustration from "Pressure" to "$T_c$" indicates the control mechanism of superconductivity by pressure. (d) Pressure-dependent normalized superconducting quasiparticle density of states (QPDOS) at 0.1 K (equivalent to tunnel conductance spectrum). $\Delta_\sigma$ and $\Delta_\pi$ refer to the two superconducting gaps. (e) Phonon-assisted electronic spectral function at $T_c$ = 12.2 K.

To further examine the structure's quantum nuclear effect and temperature shift to its superconductivity, we calculate its phonon-assisted electronic spectral function at $T_c$ = 12.2 K [Fig. 3(e)] by the special displacement method [50,51]. The insignificant broadening of band dispersion demonstrates their negligible effect to the superconductivity. We also examine the isotope effect to the superconductivity of $MgB_3$. The variation of $T_c$ is very small for the $^{26}$MgB$_3$ ($\Delta T_c = -0.04$ K) and Mg$^{11}$B$_3$ ($\Delta T_c = -0.09$ K), analogous to MgB$_2$ [75].

### D. Topological and nodal superconducting nature

The kagome MgB$_3$ shows both the topology properties and superconductivity, while whether is it a topological superconductor which can play a key role in the future quantum devices, quantum computation and quantum information? We further examine its superconductor nature by a recently established method of symmetry indicators (SIs) [56]. This method enables efficient diagnosis of topological properties by examining irreducible representations (irreps) of space groups, and has been extended to superconductors [59-61,76,77].

We obtain the symmetry-based diagnosis of topological superconductor for MgB$_3$ including the spin-orbital coupling from the QEIRREPS [57] and Topological Supercon [58] tools, as shown in Table I. Each superconductor for each pairing symmetry (i.e., the symmetry property of Cooper pairs) falls into one of the following four cases: (I) representation-enforced nodal superconductor (NSC); (II) symmetry-diagnosable topological superconductor (TSC) or topological NSC (TNSC); (III) topologically trivial or not symmetry-diagnosable TSC; (IV) silent for the trivial pairing due to no band labels can be defined at any high-symmetry momenta [60].

The point group of kagome MgB$_3$ is $D_{6h}$ which contains eight representations (see Table I). We find that the MgB$_3$ superconductor with the first six pairing symmetries is a representation-enforced NSC (i.e., case I), in which compatibility relations (CRs, or named symmetry constrains) along various node-line paths are violated. It is also found that the shapes of these nodes are point-type ($P$) for *odd*-parity cases and line-type ($L$) for *even*-parity cases, respectively. For instance, the case of $B_{1u}$ pairing is a NSC with point-type node where CRs are broken along the A-L path. In contrast, all CRs are satisfied for the $A_{1u}$ pairing symmetry. Thus, we can diagnose the topology by SIs. A fully gapped TSC or TNSC phase (case II) may be realized for the $A_{1u}$ pairing symmetry, which belongs to the entry (0, 0, 0, 0, 1, 1, 0, 1, 11, 7) of SIs. This nontrivial (i.e., nonzero array) SI of $A_{1u}$ pairing indicates its topologically nontrivial nature. While the trivial $A_{1g}$ pairing is case IV and no band labels can be defined at any high-symmetry momenta [60].

It is worth noting that MgB$_3$ shows the intrinsic nodal superconductivities in most of the pairing symmetries shown in Table I. Differently, some transition metal dichalcogenide monolayers (e.g., NbSe$_2$ and TaS$_2$) can become the nodal superconductors when they are driven by a high enough external in-plane magnetic field [78,79]. It is also recently unveiled that their current-phase relation in a Josephson junction geometry is obviously dependent on the momentum transverse to the current direction and displays a distinctive $4\pi$ periodicity [80]. In addition, unlike that for some monolayers such as graphene, a uniaxial stretch could indue a pseudomagnetic field which will influence its electronic and other properties [81], here the strain applied on the MgB$_3$ is the hydrostatic pressure which gives rise to the almost uniform deformation and negligible pseudomagnetic field effect in this bulk.

TABLE I. Results of symmetry-based diagnosis for Kagome topological superconductor MgB$_3$. The first and second column lists the point group (PG) and pairing symmetries of kagome MgB$_3$. The third one indicates case I-IV for each pairing symmetry. Symbols [$P$] and [$L$] for case I specify the shape of the nodes, point- and line-type, respectively. The fourth column is the paths where compatibility relations (CRs, or named symmetry constrains) are violated for case I and the entry of symmetry indicators for case II. The fifth column represents which types of topologies should appear.

| PG | Pairing | Case | Nodes | Topology |
|---|---|---|---|---|
| $D_{6h}$ | $B_{1u}$ | I [$P$] | A-L | NSC |
| | $B_{1g}$ | I [$L$] | A-H, A-L, M-L | NSC |
| | $B_{2u}$ | I [$P$] | A-H, K-H | NSC |
| | $B_{2g}$ | I [$L$] | A-H, A-L, M-L, K-H | NSC |
| | $A_{2u}$ | I [$P$] | M-L, K-H | NSC |
| | $A_{2g}$ | I [$L$] | A-H, A-L, M-L, K-H | NSC |
| | $A_{1u}$ | II | (0,0,0,0,1,1,0,1,11,7) | TSC or TNSC |
| | $A_{1g}$ | IV | … | … |

### E. STM and ARPES pictures

To show more observable physical properties of the two-band MgB$_3$ superconductor which can be compared with experiments, we simulate its STM and ARPES pictures in Fig. 4, respectively. The kagome lattice of B layer is easily observed in the bright region of the STM picture, and the black holes refer to the Mg atoms [Fig. 4 (a)]. Based on a surface Green's function model shown in Fig. 4(b) [41,53,55,82,83], we also obtain the ARPES picture (i.e., surface band) of the MgB$_3$ (001) surface [Figs. 4(c)-4(e)], whose work function Φ is 4.9 eV smaller than that of MgB$_2$ superconductor (6.0 eV) [see Supplemental Material [71] for details]. Figure 4(c) shows the surface system's whole surface states and bulk states. The surface states of the outermost monolayer 1L are drawn in Fig. 4(d), and the bulk states become clearer with the increase of detection depth D such as to the third layer 3L [Fig. 4 (e)].

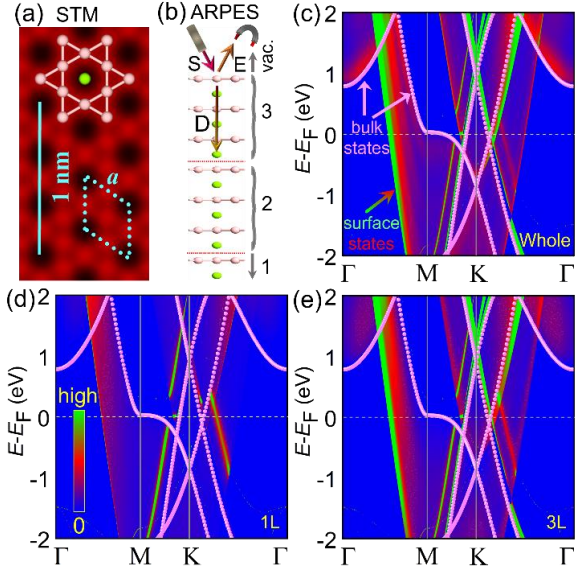

FIG. 4. (a) Simulated STM picture of MgB$_3$ (001) surface. (b) Schematic of angle-resolved photoemission spectroscopy (ARPES) set by the surface Green's function model, which includes the periodic bulk region 1, screening region 2, surface region 3, and vacuum (vac. = 15 Å). D is the detection depth. S and E refer to the photon source and emitted light, respectively. ARPES (i.e., surface band) of the whole system (c), and with detection depths to the outermost monolayer 1L (d) and the third layer 3L (e). Band of bulk is also shown in (c)-(e) with pink balls.

## IV. CONCLUSIONS

In conclusion, we find a new type-II two-band MgB$_3$ superconductor with a simple kagome lattice structure, light atomic weights, and a high intrinsic $T_c$ = 12.2 K (*highest* among the known kagome-type superconductors). Its exotic physical properties are revealed by *ab initio* calculations, including the van Hove singularity, flat-band, multiple Dirac points, and topologically nontrivial properties. It is a s-wave and anisotropic superconductor based on its anisotropic EPC $\lambda_{n\mathbf{k}}$ and superconducting gap $\Delta_{n\mathbf{k}}$. The lattice strain can significantly tune its $T_c$ (e.g., up to 15.4 K), EPC $\lambda$, and $\Delta$ in a large range, and the quantum phase transition from normal metallic to superconducting and unstable states can be driven by external stress. The topological and nodal superconducting nature of MgB$_3$ is also unveiled by a recently developed symmetry indicators method. Its pressure-dependent normalized superconducting quasiparticle density of states (equivalent to experimental tunneling conductance), STM, and ARPES pictures are systematically simulated to unveil its remarkable physical properties. Our results demonstrate that the kagome MgB$_3$ can become one important superconducting and topological material leveraged for new low-power devices. This work also paves a way to explore more MgB$_3$-type kagome superconductors and other physical properties (e.g., topology) of kagome systems with XY$_3$ formula.

## ACKNOWLEDGMENTS

Work in China was supported by the National Natural Science Foundation of China (Grant Nos. 12274117, 11888101, 61835013, 12174461, 12234012, 62274066, and 62275074), the National Key R&D Program of China (Grant Nos. 2021YFA1400900, 2021YFA0718300, 2021YFA1402100), the Science Foundation for the Excellent Youth Scholars of Henan Province (Grant No. 202300410226), the Young Top-notch Talents Project of Henan Province (2021 year), the Scientific and Technological Innovation Program of Henan Province's Universities (Grant No. 20HASTIT026), Space Application System of China Manned Space Program, and the HPCC of HNU. Work in the U.S. was supported by the U.S. DOE, Basic Energy Science (Grant No. DE-FG02-05ER46237). We thank M. Kawamura and S. Ono at the University of Tokyo, M. Zacharias at Cyprus University of Technology, and K. Jin at Institute of Physics of CAS for helpful discussions.